\documentclass{emulateapj}
\usepackage{amsmath}

\def\bluered{The blue dotted (red dashed) curve shows the results from the analysis of only LC1 (LC2). }
\def\kms{\,{\rm km\,s^{-1}}}
\def\cm{\,{\rm cm}}

\def\rE{\,R_{\rm E}}
\def\mmass{\left<M/M_\sun\right>}
\def\deg{^\circ}
\def\dos{D_{\rm OS}}
\def\RV{R_{\rm V}}
\def\projArea{\pi \RV^2\cos i}

\begin{document}
\title{Microlensing Evidence That A Type 1 Quasar Is Viewed Face On}

\author{Shawn Poindexter\altaffilmark{1}, Christopher S. Kochanek\altaffilmark{1}}
\altaffiltext{1}{
Department of Astronomy and Center for Cosmology and AstroParticle Physics,
The Ohio State University, 
140 W 18th Ave, Columbus, OH 43210, USA,
(sdp,ckochanek)@astronomy.ohio-state.edu}

\begin{abstract}

Using a microlensing analysis of 11-years of OGLE V-band photometry of
the four image gravitational lens Q2237+0305, we measure the
inclination $i$ of the accretion disk to be $\cos i > 0.66$ at 68\% confidence.
Very edge on ($\cos i < 0.39$) solutions are ruled out at 95\% confidence.
We measure the V-band radius of the accretion disk,
defined by the radius where the temperature matches the monitoring band
photon emission, to be
$\RV = 5.8^{+3.8}_{-2.3}\times 10^{15}\cm$
assuming a simple thin disk model and including the uncertainties in its inclination.
The projected radiating area of the disk remains too large to be consistent with the
observed flux for a $T\propto R^{-3/4}$ thin disk temperature profile.
There is no strong correlation between the direction of motion
(peculiar velocity) of the lens galaxy and the orientation of the disk.

\end{abstract}

\keywords{
gravitational lensing ---
methods: numerical ---
quasars: general ---
quasars: individual (Q2237+0305) ---
}

\section{Introduction} \label{sec:intro}

In the AGN unification model \citep[e.g.][]{Antonucci93,Urry95},
orientation differences among intrinsically similar objects are thought to
account for many of the different observational properties of AGN.
In particular, a dusty ``torus'' may frequently obscure the central engine from
direct observation when viewed edge-on.
The bright, Type 1 broad line quasars are thought to be viewed
mostly face-on, so that the accretion disk is not obscured by material
in the equatorial plane, and Type 2 narrow line quasars are viewed closer to edge-on
through the obscuring material.
Thus, if we could resolve the disk of a Type 1 quasar, we would expect it to be closer
to face-on than edge-on.
Unfortunately, familiar methods cannot resolve accretion disks, so we have only
indirect measures of AGN disk orientation.
For example, the projected axes of radio jets
\citep{Blandford79} and ionization cones \citep{Elvis00} both support this picture.
There are, however, no
actual measurements of disk orientation.

While quasar accretion disks are too small to be resolved by
direct imaging, gravitational microlensing provides a natural telescope to
study the structure of quasar accretion disks and the properties
of cosmologically distant lens galaxies where we see multiple images of
background quasars \citep[see][]{Wambsganss06}.
In addition to the mean potential of the lens galaxy, each image
is also magnified by the microlensing effects of the nearby stars.
Since the observer, the lens galaxy and its stars,
and the quasar are all moving, microlensing is observed as uncorrelated time
variability in each of the quasar images.
The amplitudes of these variations depend on the
structure of the accretion disk and the properties of the lens galaxy.

Quasar microlensing is most sensitive to the projected area of the
accretion disk relative to the source plane Einstein radius,
\begin{eqnarray}
\rE &=& D_{\rm OS} \sqrt{\frac{4G\left<M\right>}{c^2}\frac{D_{\rm LS}}{D_{\rm OL}D_{\rm OS}}}
\nonumber \\
&=& 1.8\times 10^{17} \left(\frac{\left<M\right>}{M_\Sun}\right)^{1/2}\cm,
\end{eqnarray}
where $G$ is the gravitational constant, $c$ is the speed of light,
$\left<M\right>$ is the mean stellar mass of the stars,
$D_{\rm LS}, D_{\rm OL}$, and $D_{\rm OS}$ are the angular diameter distances between
the lens-source, observer-lens, and observer-source respectively,
and we have used the lens and source redshifts for Q2237+0305
\citep[$z_{\rm l}=0.0394\,,z_{\rm s}=1.685$,][Q2237 hereafter]{Huchra85}.
The smaller the accretion disk, the higher the variability amplitude from microlensing.
In general, the emission profile is difficult to determine, as models
having similar half light radii show similar microlensing variability
\citep{Mortonson05}.
There has been little examination of other structural parameters of disks, except
for \citet{Congdon07}, who demonstrated in simulations that the microlensing signal
is sensitive to the ellipticity and orientation of the accretion disk.
If detectable in practice, measuring the apparent ellipticities of accretion
disks provides an important test of AGN unification models and opens the
possibility of examining the complex effects of relativity on the
apparent surface brightness of the disk \citep[e.g.][]{Hubeny01}.

Measurements of quasar disk sizes using microlensing are now common.
Recent efforts have studied individual sizes \citep[e.g.][]{Morgan08a},
the relationships between size and wavelength
\citep{Anguita08,Bate08,Eigenbrod08b,Poindexter08,Floyd09,Mosquera09},
size and black hole mass \citep{Morgan10}, and the sizes of thermal and
non-thermal emission regions \citep{Pooley07,Morgan08b,Chartas09,Dai09}.
All these studies used circular accretion disks and static
magnification patterns that neglect the random motion of stars in the lens galaxy.

\begin{figure*}
\plotone{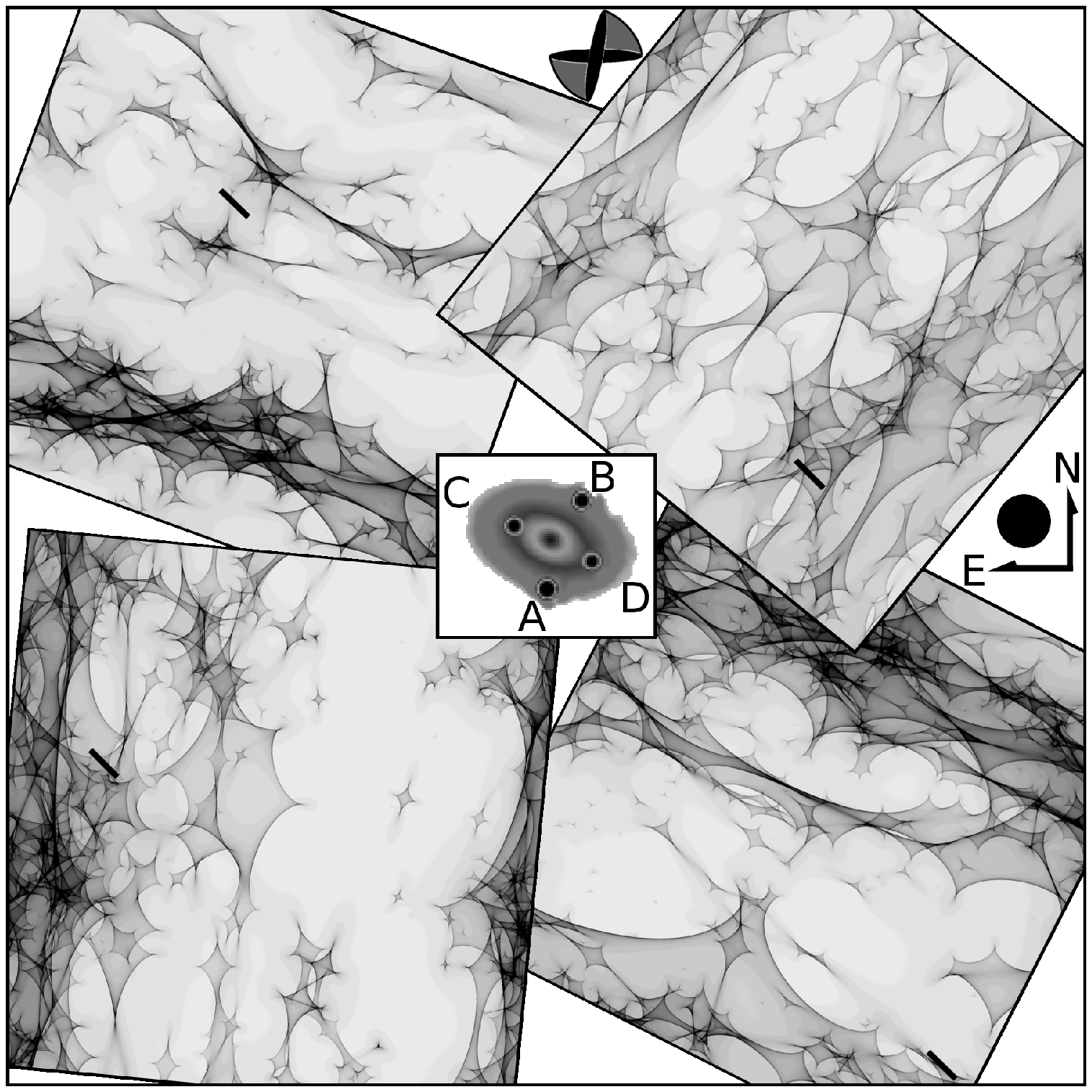}
\caption{Example of a trial source trajectory (dark line segments) superposed on
instantaneous point-source magnification patterns for $\mmass = 0.3$.
Darker shades indicate higher magnification.
An HST H-band image in the center labels the images and the corresponding
magnification patterns.
Each pattern is rotated to have the correct orientation relative to the lens.
This particular LC2 trial has an effective lens-plane velocity of
$\sim 600\kms$ Northeast.
The ``bow tie'' (top) exhibits the two peaks in the major-axis position angle
distribution together with the shaded region it approximately represents
the 68\% confidence region.
The solid disk (right) has a radius of $10^{17}\cm$.}
\label{fig:patterns}
\end{figure*}

Determining the shape and orientation of a disk depends on the correlations
between the anisotropic structure of the disk and the anisotropic structure in
the magnification patterns created by the shear (tidal gravity) local to each image
(see Figure \ref{fig:patterns}, \citealt{Congdon07}).
Existing microlensing studies cannot safely explore these issues because they
neglect the motions of the stars in the lens galaxy and use ``static''
magnification patterns.
Since the stellar velocity dispersions of lens galaxies are comparable to the peculiar
velocities of galaxies, the patterns change on the same time scale as the
source traverses them.
Ignoring these stellar motions will overestimate
the coherence of the magnification patterns \citep[see][]{Wyithe00,Kochanek07}
and likely render estimates of disk shapes unrealistic.
With a few exceptions that do not focus on disk structure (see Paper I and
\citet{Wyithe99}),
analyses of microlensing data have used static magnification patterns
because of the computational challenges.
In Paper I \citep{Poindexter09}, we solved these computational problems and examined
the peculiar velocity of the lens galaxy of Q2237
and the mean mass of its stars.
In this paper, we measure the size, inclination,
and position angle of the accretion disk in Q2237.
In \S\ref{sec:data} we describe the data set used for our microlensing analysis,
our disk model, and outline our overall approach.
Our results are presented in \S\ref{sec:results}, and a discussion follows in 
\S\ref{sec:discussion}.

\section{Data and Methods} \label{sec:data}

We analyze the nearly 11 years of Optical Gravitational Lensing Experiment (OGLE)
V-band photometric monitoring data for Q2237 \citep{Udalski06}.
To speed our analysis and as a cross check on the results,
we divided the data into two separate light curves.
The first light curve (LC1 hereafter) ranges from JD 2,450,663 to
JD 2,452,621 and consists of 100 epochs.
The second light curve (LC2 hereafter) has 230 epochs from JD 2,452,763 to
JD 2,454,602.  Each light curve covers just over 5 years.
We broaden the OGLE uncertainties by our estimate of the
systematic uncertainties in the photometry of 0.02, 0.03, 0.04, and 0.05
magnitudes in quadrature \citep{Poindexter09}.

We analyze these light curves using the Bayesian Monte Carlo method
of \citet{Kochanek04}, expanded to include motions of stars as detailed in Paper I.
For each epoch of the light curve we generate a magnification pattern including
the random motion of the stars.  We used fixed mean masses of
$\left<M\right> = 0.01, 0.03, 0.1, 0.3, 1, 3,$ and $10~M_\sun$, and
a mass function of $dN/dM \propto M^{-1.3}$ with a dynamic range
$M_{\rm max}/M_{\rm min} = 50$ based on \citet{Gould00}.
In Paper I we find that the best fit to the data is for $\left<M\right> = 0.3M_\odot$.
The stars are assigned a random velocity in each coordinate from a Gaussian
distribution with $\sigma = 170\kms$ based on the observed velocity dispersion
(van de Ven, G. 2009, personal communication, \citet{Trott08}).
The orbit of Earth (parallax effect) and the rotation of the lens galaxy
are both included in the simulation.
We convolve the patterns with the disk models described in \S\ref{sec:data}.
We draw the bulk velocities of the observer, lens galaxy, and source from a
Gaussian of dispersion $\sigma = 1000\kms$ on the lens plane in each coordinate.
We later reweight the results to a more compact velocity prior based on the projection
of the CMB dipole velocity \citep{Hinshaw09} onto the lens plane, which is small
$(-50,-23)\kms$, and the (1D) peculiar velocity dispersions of the lens and
source of $327\kms$ and $230\kms$ \citep[estimated from][]{Tinker09}, respectively.
Because of the low lens redshift and the small projected dipole, the peculiar velocity
of the lens is by far the most important factor (see Paper I).
We then randomly draw light curves for each image and fit them to the data.
Bayes theorem is used to combine the goodness of fit for the trials
as measured by a $\chi^2$ statistic, into probability distributions for each
variable of interest.  These procedures are described in detail in
\citet{Kochanek04} and Paper I.

\subsection{Accretion Disk Model}
\label{sec:diskmodel}

We employ a generic thin disk model for which the surface temperature
scales as $T \propto R^{-3/4}$ with radius R \citep{Shakura73}.
The microlensing signal is primarily sensitive to the half-light radius of the
disk \citep{Mortonson05}, which controls the effective smoothing area of the
disk, and, to date, studies have been unable to distinguish differing radial
profiles \citep[e.g.][]{Kochanek04}.
A more realistic disk model would include a central hole whose size depends on
the last stable orbit and general relativistic effects modify the underlying
intensity profiles \citep[e.g.][]{Hubeny01}.
We have not presently pursued these features because they add many model parameters, and 
because
it is unclear whether the temperature profile of the thin disk model is correct.
In particular, microlensing studies suggest the need for a flatter temperature
profile \citep[e.g.][]{Pooley07,Morgan10,Poindexter08,Eigenbrod08b}
in order to reconcile the microlensing sizes with the observed optical fluxes.
Such changes in the temperature profile have also been suggested to explain
the deviations of the observed spectral slope from the thin disk emission
(see the reviews by \cite{Koratkar99} and \cite{Blaes04}).

Thus, we assume that the face-on ($\cos i = 1$) surface brightness of the disk is
\begin{equation}
f_\nu = \frac{2hc}{\lambda^3_{\rm rest}} \big[ \exp{(R/R_{\lambda})^{3/4}}-1 \big]^{-1},
\end{equation}
where
\begin{eqnarray}
R_{\lambda} &=& \left[ {45 G \lambda_{rest}^4 M_{\rm BH} \dot{M} \over 16 \pi^6 hc^2} \right]^{1/3}
\nonumber \\
&=& 9.7 \times 10^{15} \left( {\lambda_{rest} \over {\rm \micron}} \right)^{4/3}
\nonumber \\ && \times
\left( {M_{\rm BH} \over 10^9 {\rm M_{\sun}}} \right)^{2/3}
\left({L \over \eta L_E} \right)^{1/3}\cm
\label{eqn:rlambda}
\end{eqnarray}
corresponds to the radius where
the disk temperature equals the photon energy, $kT=hc/\lambda_{\rm rest}$,
$M_{\rm BH}$ is the black hole mass, $\dot M$ is the accretion rate,
$L/L_{\rm E}$ is the luminosity relative to Eddington luminosity, and
$\eta = L/(\dot M c^2)$ is the radiative efficiency of the accretion disk.
The half light radius of the disk is
$R_{1/2} = 2.44 R_\lambda$. We now simply treat the disk as an infinitely 
thin disk viewed at inclination angles $i$, selected from a uniform distribution
in $\cos i = 0.2, 0.4, 0.6, 0.8$, and $1.0$ (face-on).
We include no relationship between the surface brightness and the viewing
angle other than this simple projection effect in this first exploration of the problem.
For full relativistic disk models, there are many complexities such as Doppler
shifts, redshifts, and bending of ray trajectories \citep[e.g.][]{Agol97},
but the bulk of the optical flux comes from relatively large radii,
$\RV/R_{\rm g} \sim 30$ for $R_{\rm g} = GM_{\rm BH}/c^2 = 2\times10^{14}\cm$,
where these effects are less important.
Here we use \citet{Morgan10}'s estimated black hole mass of
$1.3\times 10^9M_\sun$ found by applying the virial relation of
\citet{Vestergaard06} to the \ion{C}{4} line width measurement from \citet{Yee91}.
Since the orientation of the projected disk relative to the magnification
patterns also affects the results, we considered 18 major axis position angles
from $0\deg$ to $170\deg$ in steps of $10\deg$, where the remaining angles are
covered by the reflection symmetry of the disk model.
We parametrized the size
of the disk by the projected area, $\projArea$, as this is likely
to have less correlation with the inclination angle because it keeps the
projected area of the disk constant.

We must also worry about whether all the observed emission arises directly
from the accretion disk.  The light curves can be contaminated by broad
line emission on much larger scales or some of the emission from the disk
can be scattered on larger scales \citep[see][]{Dai09,Morgan08b,Morgan10}.
At V-band, the contamination from broad line emission, mainly
Fe pseudo-continuum emission and \ion{C}{3}$]$~$\lambda1909$,
contributes of order $20\%$ of the flux in the spectral models of \citet{Eigenbrod08a}.
To examine the effect of this dilution we ran models with $0\%$, $20\%$,
and $40\%$ contamination by emission on large, un-microlensed scales for
the $\left<M\right> = 0.3M_\odot$ case.
Adding an unmicrolensed contamination fraction, $f$, in the V-band also
decreases the flux size estimate by $(1-f)^{1/2}$ unless the contamination
is due to scattering of the disk emission on large scales
\citep[see the discussion in][]{Morgan10}.

\section{Results} \label{sec:results}

We estimate the projected area of the accretion disk $\projArea$,
the deprojected radius $\RV$, the disk inclination $i$, and
the major axis position angle.
We also do a limited set of tests with different amounts of unmicrolensed
flux that may influence our analysis.
We quote the results from the combined analysis of LC1 and LC2,
but also show the results from the independent analyses of LC1 and LC2.
Since the results are always mutually consistent,
we only report quantitative results for the combined analysis.

\begin{figure}
\epsfxsize=3.5truein
\epsfbox{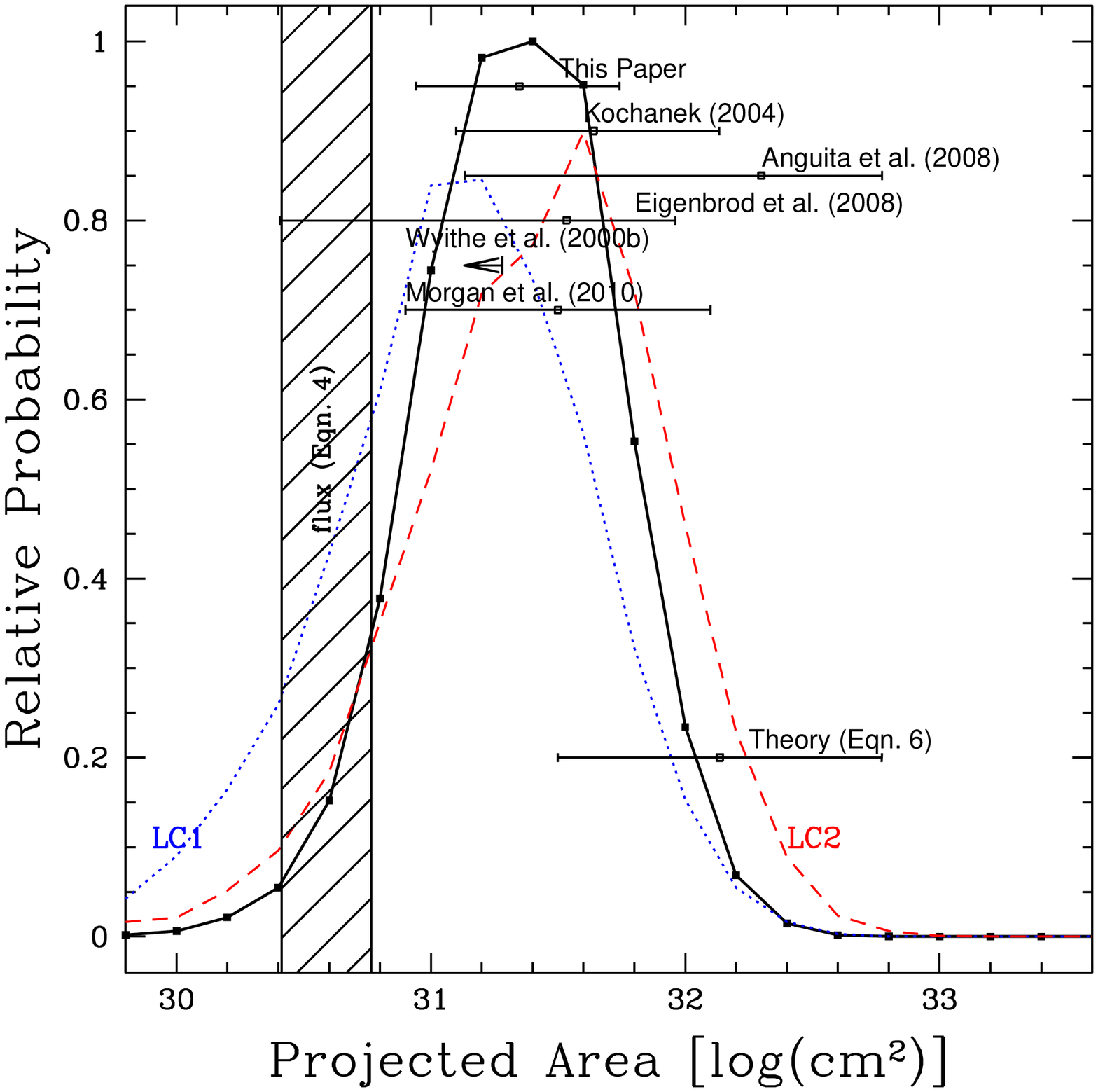}
\caption{The projected area $\projArea$ distribution of the accretion disk. \bluered
The black curve is the combined result from both LC1 and LC2.
The horizontal bars compare 68\% confidence regions from earlier studies assuming
$\left<M\right>=0.3M_\odot$ for estimates which depended on the mean mass.
The upper limit found by \citet{Wyithe00b} is at 99\% confidence.
The ``flux'' estimate is the area predicted based on the observed
flux (Equation \ref{eqn:fluxsize}) and is independent of inclination.
The ``theory'' estimate is the area thin disk theory predicts assuming $\cos i=1$,
based on the estimated black hole mass for Q2237 (Equation \ref{eqn:rlambda2}).
\label{fig:area}}
\end{figure}

\begin{figure}
\epsfxsize=3.5truein
\epsfbox{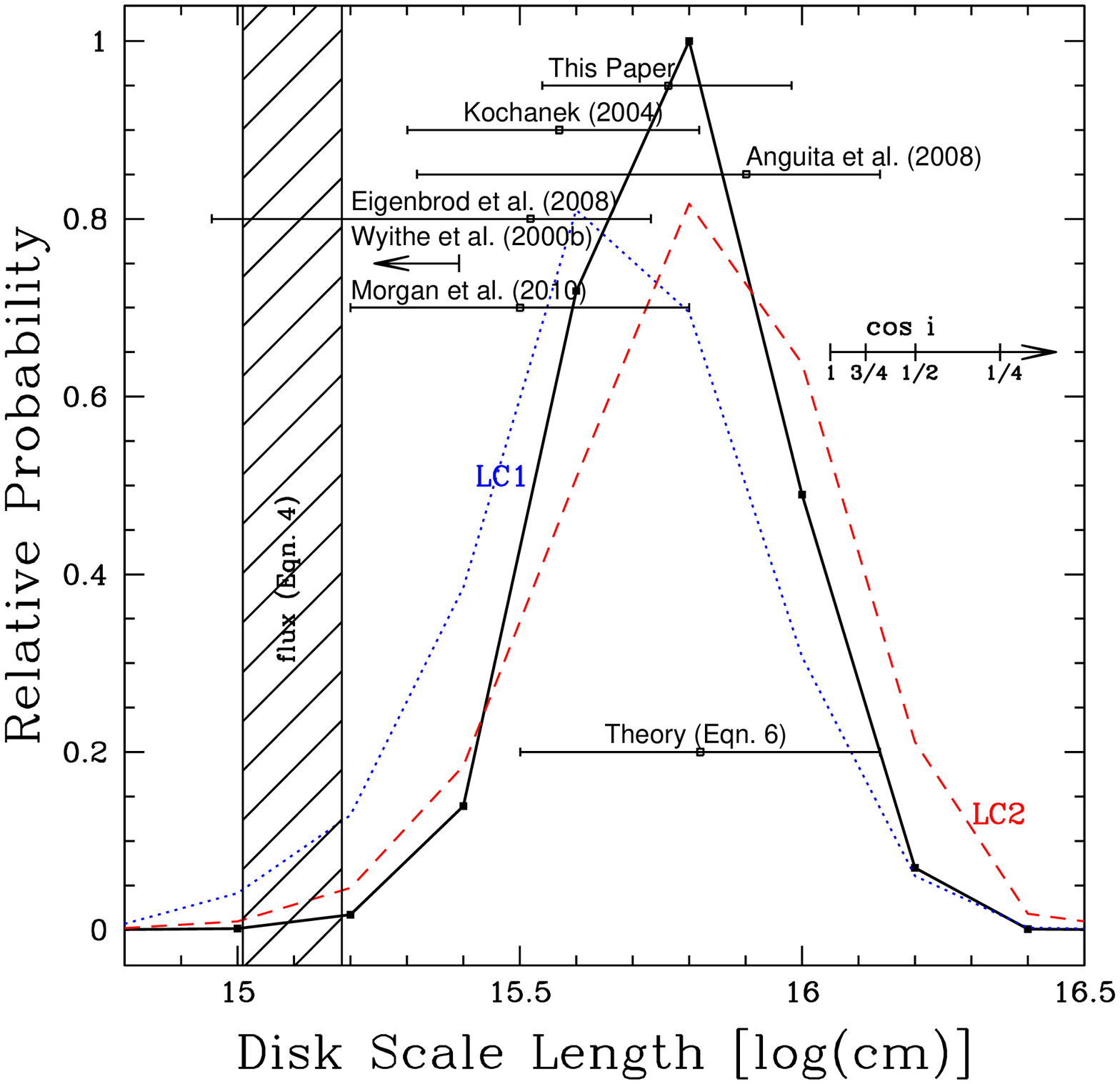}
\caption{The disk scale length $\RV$ after de-projection from the joint analysis
of LC1 and LC2.  \bluered
The earlier results are shown assuming $\cos i=1$ for these face-on models.
A scale on the right shows how they would shift to larger sizes if $\cos i < 1$.
The flux size has the same inclination dependence as the earlier results,
while the theory size is independent of inclination.
Our test with 20\% (40\%) of light being contamination emitted on
much larger scales resulted in a 26\% (55\%) smaller radius.
\label{fig:radius}}
\end{figure}

We find (Figure \ref{fig:area}) the projected V-band area of the quasar
defined by $\projArea$ to be
$8.7\times10^{30}\cm^2 < \projArea < 5.5\times10^{31}\cm^2$
($4.7\times10^{30}\cm^2 < \projArea < 9.4\times10^{31}\cm^2$)
at 68\% (95\%) confidence with a median of $\projArea = 2.2\times10^{31}\cm^2$,
where the size scale $\RV$ is defined by Equation \ref{eqn:rlambda}.
After de-projecting the area based on each trial's inclination, we find that the 
V-band radius of the accretion disk is
$3.5\times10^{15}\cm < \RV < 1.0\times10^{16}\cm$
($2.6\times10^{15}\cm < \RV < 1.4\times10^{16}\cm$)
at 68\% (95\%) confidence with a median value of $\RV = 5.8\times10^{15}\cm$
(Figure \ref{fig:radius}).
This is consistent with our earlier results in
\citet{Kochanek04} and \citet{Morgan10}, of
$\RV = 3.7^{+2.9}_{-1.7}\times10^{15}/\sqrt{\cos i}\cm$ and
$\RV = 3.2^{+3.1}_{-1.6}\times10^{15}/\sqrt{\cos i}\cm$
using the same method without dynamic patterns,
a smaller velocity prior, and shorter light curves.
Other analyses by
\citet{Wyithe00b}, \citet{Yonehara01}, \citet{Vakulik07}, \citet{Anguita08}, and
\citet{Eigenbrod08b}
have found generally consistent results of
$\RV < 2.5\times10^{15}\cm$ (at 99\% confidence),
$\RV \la 1.8\times10^{17}\cm$,
$\RV \sim 2\times10^{15}\cm$,
$\RV = 8.0^{+5.7}_{-5.9}\times 10^{15} \sqrt{M/0.3 M_\sun}\cm$, and
$\RV = 3.3^{+2.1}_{-2.4}\times10^{15}\cm$ (see Figure \ref{fig:radius}),
but using less data and with far stronger systematic assumptions.
These estimates are all for $\cos i=1$ and should scale as $1/\sqrt{\cos i}$.
If we compare projected areas (Figure \ref{fig:area}), then the comparisons are
(to 1st order) independent of the inclination angle, while if we compare disk scale
lengths, there will be an inclination angle dependence.
While generally consistent with our results, our calculations use more data and are considerably
more realistic, making it difficult to evaluate differences, especially the uncertainties.

We can compare our measurement to those predicted by the observed fluxes or
thin disk theory.
If we assume only the $T\propto R^{-3/4}$ temperature profile of a thin disk model
\citep{Shakura73}, where we can ignore the inner disk edge for these wavelengths, then the
observed flux constrains the disk size by matching the integrated flux from a disk
with the emission profile of Equation \ref{eqn:rlambda} to the observed flux.
We estimate that the magnification corrected
I-band flux is $I = 18.03\pm0.44$ mag, and this corresponds to a disk radius of
\begin{eqnarray}
\RV = && 1.7\times 10^{15} \frac{1}{\sqrt{\cos i}} \left(\frac{\dos}{r_{\rm H}}\right)
\nonumber \\
&& \times \left(\frac{\lambda_{\rm I,obs}}{\mu \rm m}\right)^{3/2} 10^{-0.2(I-19)} h^{-1}\cm,
\label{eqn:fluxsize}
\end{eqnarray}
where $\dos/r_{\rm H}$ is the angular diameter distance to the quasar relative to the
Hubble radius, $r_{\rm H}$ \citep[see][]{Morgan10}.
This gives $\projArea = 3.9^{+1.9}_{-1.3}\times10^{30}\cm$.
Assuming our best fit inclination, $\cos i = 0.8$,
$\RV = 1.25^{+0.28}_{-0.23}\times10^{15}\cm$.

\citet{Agol09} estimate that the bolometric luminosity of the quasar is
$L_{Agol} = 4 \times 10^{46}$~ergs~s$^{-1}$, which corresponds to a Eddington factor of
\begin{equation}
{ L \over L_E } = { 1 \over 3 } { L_{bol} \over L_{Agol}}
{ 10^9 M_\odot \over M_{BH}},
\label{eqn:lle}
\end{equation}
that is typical of luminous quasars \citep{Kollmeier06}.
If we use this to replace the $L/L_E$ factor in Eqn. \ref{eqn:rlambda}, the size estimate becomes
\begin{equation}
R_\lambda = 6.6 \times 10^{15}
\left( { \lambda_{rest} \over \mu\hbox{m} } \right)^{4/3}
\left( { M_{BH} \over 10^9 M_\odot } { L_{bol} \over L_{Agol}} { 1 \over \eta } \right)^{1/3}\cm,
\label{eqn:rlambda2}
\end{equation}
which now depends relatively weakly on the black hole mass.
The uncertainties in these estimates are logarithmic, corresponding
to $0.4$ and $0.3$~dex respectively for Eqns. \ref{eqn:rlambda} and \ref{eqn:rlambda2} if the
masses, Eddington factors, luminosities and efficiencies are
viewed as being uncertain by a factor of $3$.
We find that the disk is large compared to the estimate based on the observed flux,
although this is modestly reduced by the inclination, and small compared to a thin disk
radiating close to Eddington with $\eta = 10\%$ efficiency.  This is a discrepancy common to all
microlensing estimates at present \citep[see][]{Pooley07,Morgan10}.

\begin{figure}
\epsfxsize=3.5truein
\epsfbox{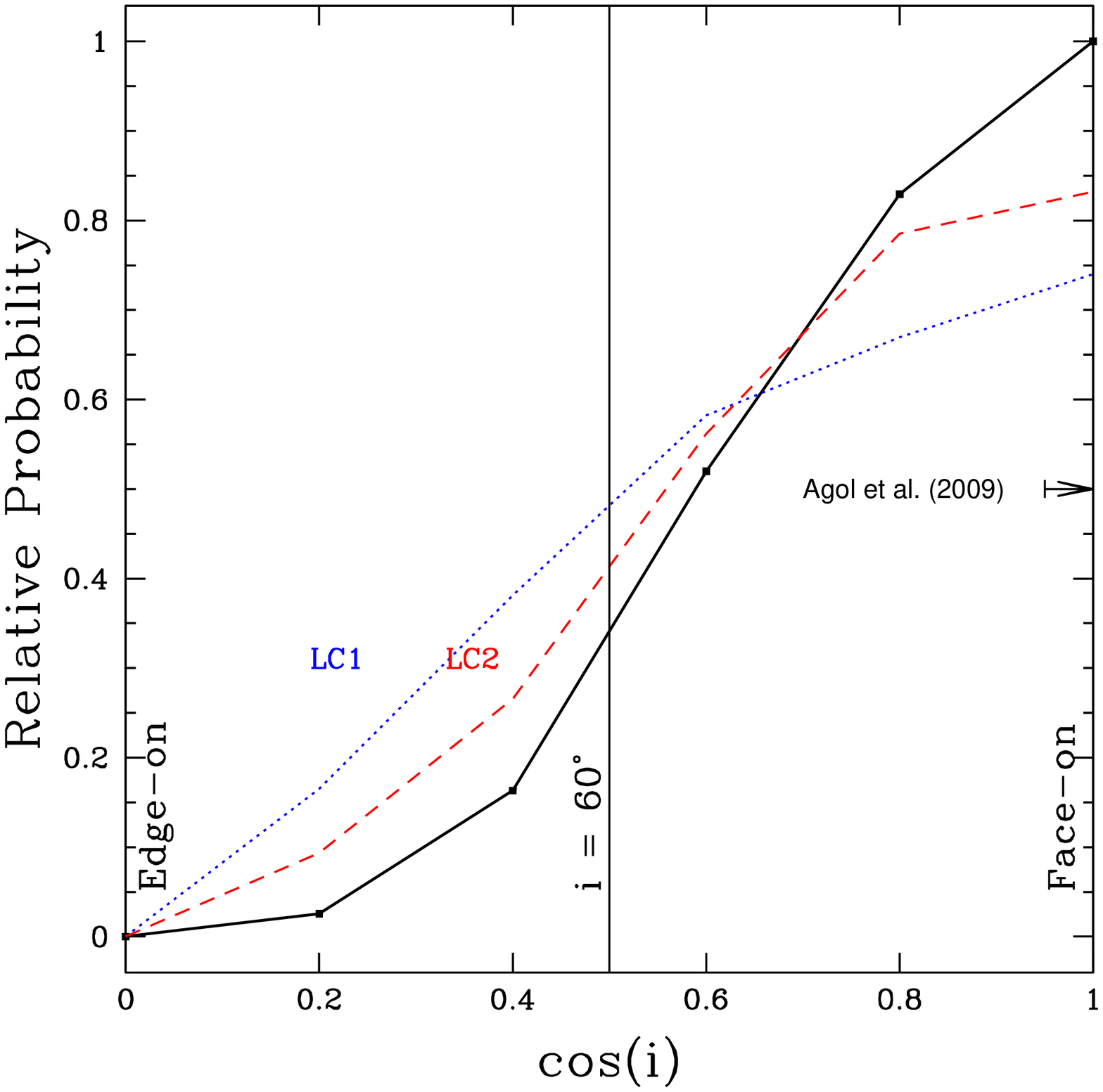}
\caption{Cosine of the disk inclination ($\cos i = 1.0$ is face-on). \bluered
The probability at $\cos i = 0$ was defined to be 0.
The \citet{Agol09} estimate is based on models of the mid-IR SED and,
according to the authors, should not be interpreted quantitatively.
\label{fig:cosi}}
\end{figure}

\begin{figure}
\epsfxsize=3.5truein
\epsfbox{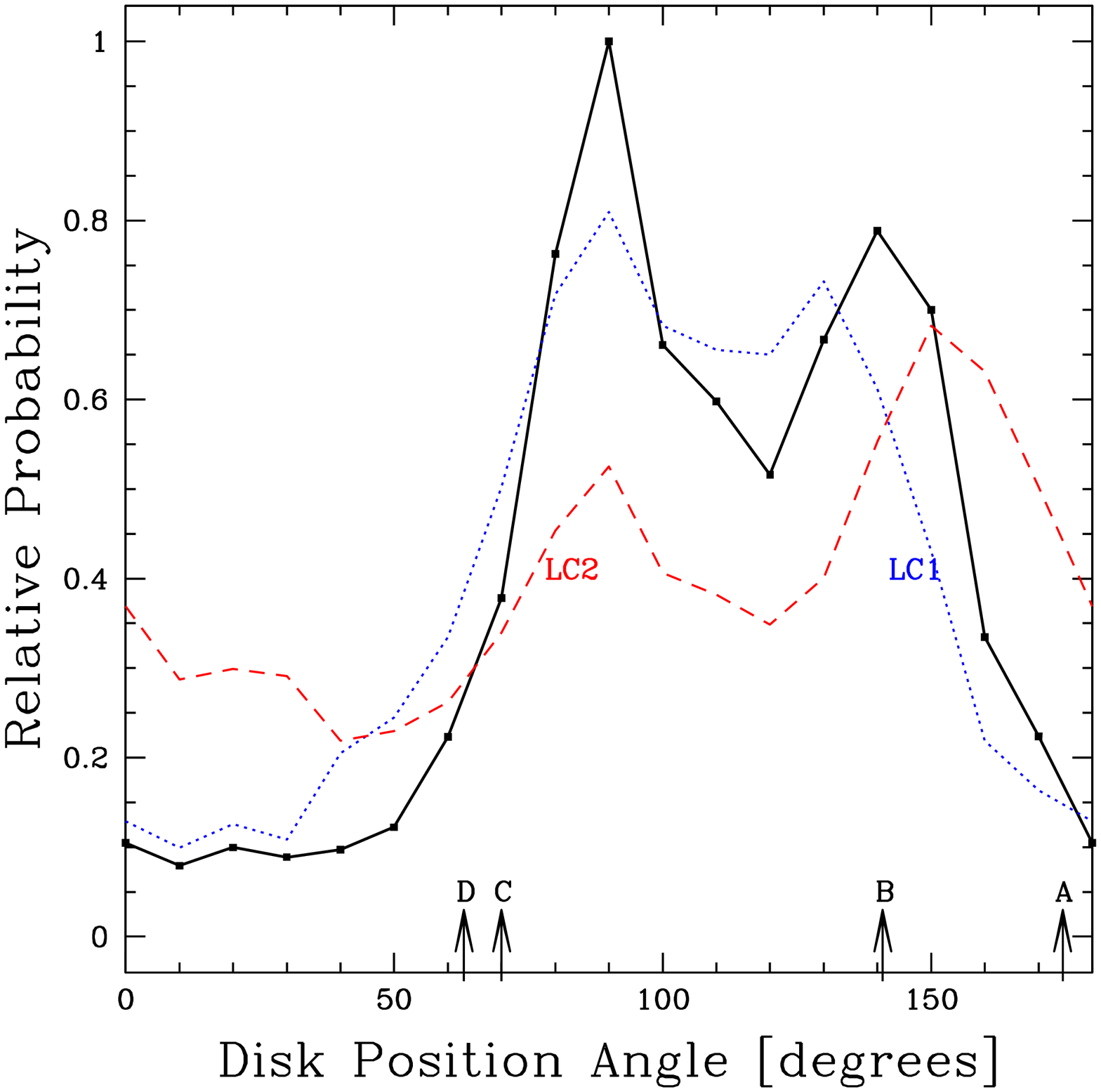}
\caption{Major axis position angle (North through East) of the accretion disk. \bluered
The shear position angles for each image are indicated by the label arrows.
The preferred orientation of the major axis is
to be parallel to the shear of images A and B but perpendicular to C and D.
The face-on solutions are not included in these distributions because they provide
no information on the position angle.
\label{fig:pa}}
\end{figure}

\begin{figure}
\epsfxsize=3.5truein
\epsfbox{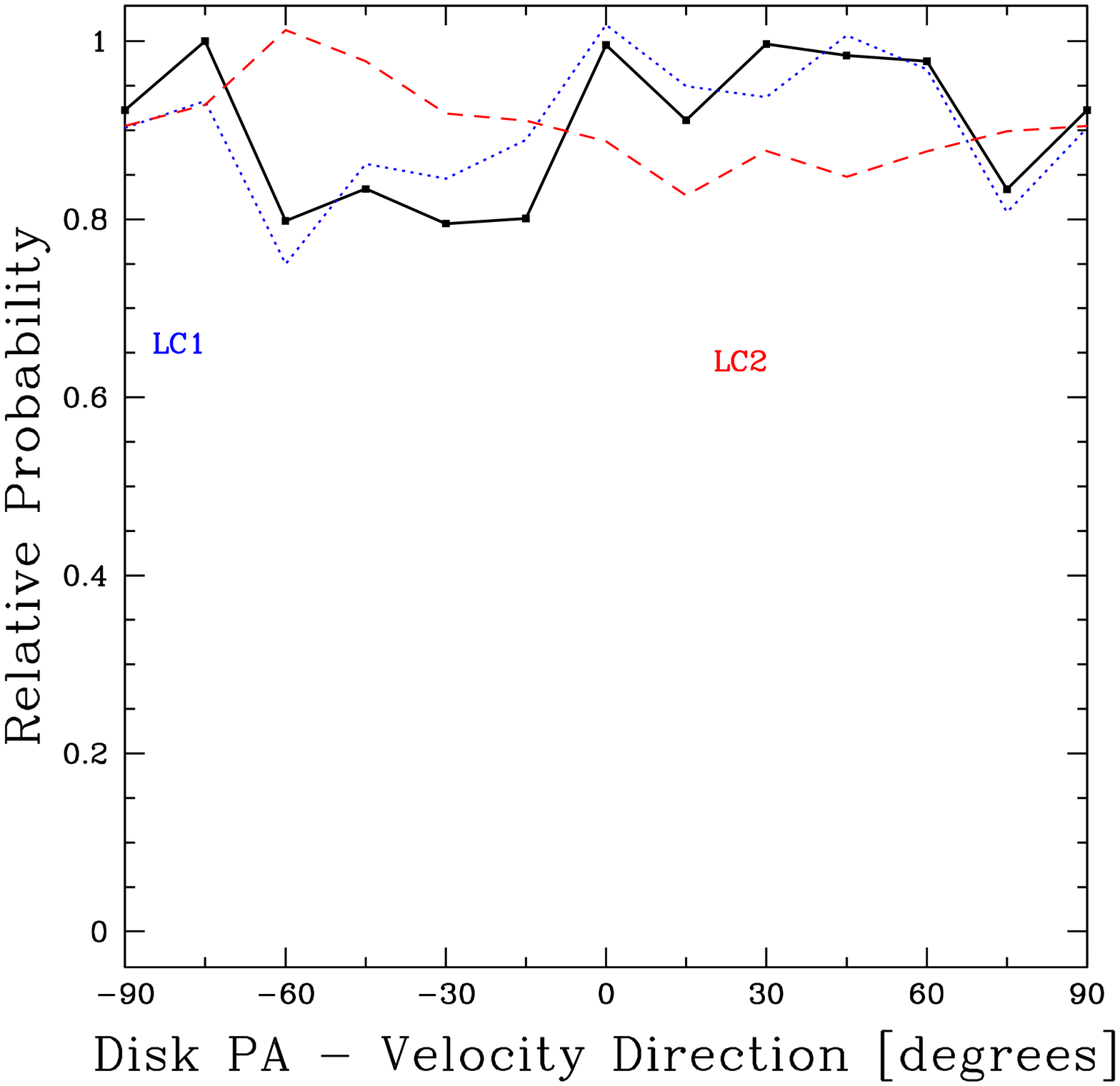}
\caption{Major axis position angle of the disk relative to the direction
of motion (N through E PAs).  Here we exclude the face-on trials since they add
no useful information to this distribution.
\label{fig:orient}}
\end{figure}

Where these size estimates are incremental improvements over earlier results from
using additional data and an improved physical model, our results for the
inclination and orientation of the disk are entirely new.
The preferred inclination is $\cos i > 0.66$ at 68\% confidence where
$\cos i = 1$ is face-on (Figure \ref{fig:cosi}). 
Such a relatively face-on inclination is consistent with the expectations of the
AGN unification model.
We can compare this to \citet{Agol09}'s model for the mid-infrared spectral
energy distribution of Q2237 using the dust torus models of \citet{Fritz06}.
While \citet{Agol09} do not trust the quantitative results, their preferred viewing angles
($i \leq 19^\circ$, Figure \ref{fig:cosi})
are consistent with our inferences for the inclination angle of the disk.

Figure \ref{fig:pa} shows our estimate of the position angle (PA) of the major
axis of the disk, and Figure \ref{fig:patterns} shows the 68\% confidence range of this
PA for comparison to the anisotropies in the magnification patterns.
We find consistent estimates of this position angle from both
LC1 and LC2 (Figure \ref{fig:pa}).
The preferred orientation is for the major axis to be roughly parallel to the shear
in image B, but perpendicular to it in images C and D.
\citet{Congdon07} found that microlensing variability is enhanced when the major
axis is aligned with the shear, so our estimated alignment helps to explain
the higher variability of images A and B, and the lower variability of D, but
not the variability of image C.
For this lens there are no other observations to which we can compare the orientation.
However, if we had observations of the quasar host galaxy, we
could compare the PA of the galaxy to that of the disk as a check on the relative
orientations of the angular momentum vectors of the accretion disk and the galaxy.
For example, \citet{Yoo05,Yoo06} used images of lensed
quasar hosts to constrain the host axis ratio and position angle in 4 lensed quasars
(to roughly $\la 20\deg$).
While both the peculiar velocity direction of the lens and the disk position
angle are constrained, we find no strong constraints on the disk orientation
relative to the direction of motion (Figure \ref{fig:orient}).
The convolved uncertainties in the two quantities are too large.

Luminosity that originates outside the accretion disk proper is a concern for size estimates
because contamination by emission on large scales dilutes the microlensing signal
and leads us to overestimate the projected area of the disk \citep[see][]{Dai09,Morgan08b}.
To examine this effect, we did a set of trials with $\mmass = 0.3$
where 0\%, 20\%, and 40\% of the source light was not microlensed.
The spectral analysis by \citet{Eigenbrod08a} suggests that the level of
contamination is 20\%.
The 0\% and 20\% cases fit equally well, while the 40\% case had a relative
probability 38\% lower.
As the dilution is increased from 0\%, more face-on disks are preferred, with
$\cos i > 0.63, 0.73,$ and $0.80$ (68\% confidence) for 0\%, 20\%, and 40\% dilution.
Adding unmicrolensed light affects the disk size both through the general dilution
and the shift towards more face-on orientations.
The de-projected radius is smaller by $\sim 26\% (55\%)$ if the contamination
is increased to 20\% (40\%) from 0\%.
These corrections are not large compared to our statistical uncertainties of
order 70\%, but they are an important physical consideration.

\bigskip
\section{Discussion} \label{sec:discussion}

By including random stellar motions in our microlensing analysis of Q2237 we find
evidence that the accretion disk of Q2237 is viewed face-on with $\cos i > 0.63$.
This lends support to the popular AGN unification model where we expect
Type 1 quasars like Q2237 to be viewed nearly face-on.
Including 20\% contamination from broad line emission on larger spatial scales
\citep[estimated from spectra][]{Eigenbrod08a} results in a stronger case for face-on solutions.
Modeling stellar motions and the inclination and position angle parameters further
reduces the systematic uncertainties of our measurements of the disk radius
compared to earlier studies by including a broader range of physical uncertainties.
As we found in Paper I for the lens velocity and mean stellar mass, the results of
the separate analyses of LC1 and LC2, the first and second temporal halves of the OGLE
light curves, produce consistent results for every parameter we considered.
While we have used a relatively simple model for the accretion disk, these results demonstrate
that disk shapes can be measured with quasar microlensing, as suggested by \citet{Congdon07}.
As data sets and computing power improve, it will be natural to try fitting more subtle disk
features such as the asymmetries from relativistic effects using relativistic models such as
\citet{Hubeny01}.

The exceptionally long OGLE light curve and the fast microlensing timescales of Q2337
made it a natural first choice for including the random stellar motions and studying the
shape of the disk.
Our expanded method is similar in computational cost to our previous efforts with static patterns,
so there is no reason not to use it generally.
The stellar motions clearly aid in reducing uncertainties in the mean mass (see Paper I)
and allow us to correctly make inclination corrections. It can also easily be extended to
take advantage of multi-wavelength data to try to constrain the temperature profile.
However, as in earlier microlensing studies \citep{Pooley07, Morgan10, Dai09}, we cannot
reconcile the basic temperature profile of a thin disk, the microlensing size estimate, and
the observed optical flux.  A disk with a $T\propto R^{-3/4}$ temperature profile normalized 
by the microlensing size estimate should be brighter than observed.  This can be solved by
altering the temperature profile \citep[see][]{Poindexter08,Morgan10}.
For example, reducing the slope from $T\propto R^{-3/4}$ to $T \propto R^{-1/2}$
would increase the flux size by a factor of $2.2$ relative to the half light radius.
This would be mildly inconsistent ($1.5\sigma$)
with the slope estimate of $-0.83 \pm 0.21$ by \citet{Eigenbrod08b}.
However such changes also call into question the basic structure of the thin disk model.
The other simple possibility is to reduce the emissivity of the disk to be well
below that of a blackbody (by the ratio of the flux/microlensing sizes squared),
but this seems less physically plausible than change in the temperature structure.

\acknowledgments
This work was supported in part by an allocation of computing time from the Ohio Supercomputer Center.
This research was supported by NSF grant AST-0708082.


\begin{thebibliography}

\bibitem[Agol(1997)]{Agol97} Agol, E.\ 1997, Ph.D.~Thesis,  

\bibitem[Agol et al.(2009)]{Agol09} Agol, E., Gogarten, S.~M., Gorjian, V., \& Kimball, A.\ 2009, \apj, 697, 1010 

\bibitem[Anguita et al.(2008)]{Anguita08} Anguita, T., Schmidt, R.~W., Turner, E.~L., Wambsganss, J., Webster, R.~L., Loomis, K.~A., Long, D., \& McMillan, R.\ 2008, \aap, 480, 327 

\bibitem[Antonucci(1993)]{Antonucci93} Antonucci, R.\ 1993, \araa, 31, 473 

\bibitem[Bate et al.(2008)]{Bate08} Bate, N.~F., Floyd, D.~J.~E., Webster, R.~L., \& Wyithe, J.~S.~B.\ 2008, \mnras, 391, 1955 

\bibitem[Blaes(2004)]{Blaes04} Blaes, O.M., 2004, in Les Houches Summer School LXXVIII (Springer: Berlin) 137

\bibitem[Blandford \& Konigl(1979)]{Blandford79} Blandford, R.~D., \& Konigl, A.\ 1979, \apj, 232, 34 

\bibitem[Chartas et al.(2009)]{Chartas09} Chartas, G., Kochanek, C.~S., Dai, X., Poindexter, S., \& Garmire, G.\ 2009, \apj, 693, 174 

\bibitem[Congdon et al.(2007)]{Congdon07} Congdon, A.~B., Keeton, C.~R., \& Osmer, S.~J.\ 2007, \mnras, 376, 263

\bibitem[Dai et al.(2009)]{Dai09} Dai, X., Kochanek, C.~S., Chartas, G., Kozlowski, S., Morgan, C.~W., Garmire, G., \& Agol, E.\ 2009, arXiv:0906.4342

\bibitem[Eigenbrod et al.(2008a)]{Eigenbrod08a} Eigenbrod, A., Courbin, F., Sluse, D., Meylan, G., \& Agol, E.\ 2008a, \aap, 480, 647 

\bibitem[Eigenbrod et al.(2008b)]{Eigenbrod08b} Eigenbrod, A., Courbin, F., Meylan, G., Agol, E., Anguita, T., Schmidt, R.~W., \& Wambsganss, J.\ 2008, \aap, 490, 933 

\bibitem[Elvis(2000)]{Elvis00} Elvis, M.\ 2000, \apj, 545, 63 

\bibitem[Floyd et al.(2009)]{Floyd09} Floyd, D.~J.~E., Bate, N.~F., \& Webster, R.~L.\ 2009, arXiv:0905.2651

\bibitem[Fritz et al.(2006)]{Fritz06} Fritz, J., Franceschini, A., \& Hatziminaoglou, E.\ 2006, \mnras, 366, 767 

\bibitem[Gil-Merino et al.(2005)]{GilMerino05} Gil-Merino, R., Wambsganss, J., Goicoechea, L.~J., \& Lewis, G.~F.\ 2005, \aap, 432, 83

\bibitem[Gould(2000)]{Gould00} Gould, A.\ 2000, \apj, 535, 928 

\bibitem[Hinshaw et al.(2009)]{Hinshaw09} Hinshaw, G., et al.\ 2009, \apjs, 180, 225 

\bibitem[Hubeny et al.(2001)]{Hubeny01} Hubeny, I., Blaes, O., Krolik, J.~H., \& Agol, E.\ 2001, \apj, 559, 680

\bibitem[Huchra et al.(1985)]{Huchra85} Huchra, J., Gorenstein, M., Kent, S., Shapiro, I., Smith, G., Horine, E., \& Perley, R.\ 1985, \aj, 90, 691

\bibitem[Kochanek(2004)]{Kochanek04} Kochanek, C.~S., 2004, \apj, 605, 58

\bibitem[Kochanek et al.(2007)]{Kochanek07} Kochanek, C.~S., Dai, X., Morgan, C., Morgan, N., Poindexter, S. \& Chartas, G., 2007, in Statistical Challenges in Modern Astronomy IV in Statistical Challenges in Modern Astronomy IV G. J. Babu and E. D. Feigelson, eds., (Astron. Soc. Pacific: San Francisco), [astro-ph/0609112]

\bibitem[Kollmeier et al.(2006)]{Kollmeier06} Kollmeier, J.A., et al. 2006, \apj, 648, 128

\bibitem[Koratkar \& Blaes(1999)]{Koratkar99} Koratkar, A \& Blaes, O. 1999, \pasp, 111, 1

\bibitem[Morgan et al.(2010)]{Morgan10} Morgan, C., Kochanek, C.S., Morgan, N.D., Falco, E.E., 2010, ArXiv e-prints, 707, arXiv:0707.0305, in prep.

\bibitem[Morgan et al.(2008)]{Morgan08a} Morgan, C.~W., Eyler, M.~E., Kochanek, C.~S., Morgan, N.~D., Falco, E.~E., Vuissoz, C., Courbin, F., \& Meylan, G.\ 2008, \apj, 676, 80

\bibitem[Morgan et al.(2008b)]{Morgan08b}  Morgan, C.~W., Kochanek, C.~S., Dai, X., Morgan, N.~D., \& Falco, E.~E.\ 2008, \apj, 689, 755

\bibitem[Mortonson et al.(2005)]{Mortonson05} Mortonson, M.~J., Schechter, P.~L., \& Wambsganss, J.\ 2005, \apj, 628, 594

\bibitem[Mosquera et al.(2009)]{Mosquera09} Mosquera, A.~M., Mu{\~n}oz, J.~A., \& Mediavilla, E.\ 2009, \apj, 691, 1292 

\bibitem[Peng et al.(2006)]{Peng06} Peng, C.~Y., Impey, C.~D., Rix, H.-W., Kochanek, C.~S., Keeton, C.~R., Falco, E.~E., Leh{\'a}r, J., \& McLeod, B.~A.\ 2006, \apj, 649, 616

\bibitem[Poindexter et al.(2008)]{Poindexter08} Poindexter, S., Morgan, N., \& Kochanek, C.~S.\ 2008, \apj, 673, 34 

\bibitem[Poindexter \& Kochanek(2009)]{Poindexter09} Poindexter, S. \& Kochanek, C.~S.\ 2009, arXiv:0910.3213

\bibitem[Pooley et al.(2007)]{Pooley07} Pooley, D., Blackburne, J.~A., Rappaport, S., \& Schechter, P.~L.  \ 2007, \apj, 661, 19

\bibitem[Shakura \& Sunyaev(1973)]{Shakura73} Shakura, N.I. \& Sunyaev, R.A., 1973, \aap, 24, 337

\bibitem[Tinker, Wetzel, \& Zehavi(2009)]{Tinker09} Tinker, J.~L. \& Wetzel, A.~R., Zehavi, I., 2009, in prep.

\bibitem[Trott et al.(2008)]{Trott08} Trott, C.~M., Treu, T., Koopmans, L.~V.~E., \& Webster, R.~L.\ 2008, arXiv:0812.0748

\bibitem[Udalski et al.(2006)]{Udalski06} Udalski, A., et al.\ 2006, Acta Astronomica, 56, 293 

\bibitem[Urry \& Padovani(1995)]{Urry95} Urry, C.~M., \& Padovani, P.\ 1995, \pasp, 107, 803 

\bibitem[Vakulik et al.(2007)]{Vakulik07} Vakulik, V.~G., Schild, R.~E., Smirnov, G.~V., Dudinov, V.~N., \& Tsvetkova, V.~S.\ 2007, \mnras, 382, 819

\bibitem[Vestergaard \& Peterson(2006)]{Vestergaard06} Vestergaard, M. \& Peterson, B.M. 2006, \apj, 641, 689

\bibitem[Wambsganss(2006)]{Wambsganss06} Wambsganss, J, 2006, in Gravitational Lensing: Strong Weak and Micro, Saas-Fee Advanced Course 33, G. Meylan, P. North, P. Jetzer, eds., (Springer: Berlin) 453, [astro-ph/0604278]

\bibitem[Wyithe et al.(1999)]{Wyithe99} Wyithe, J.~S.~B., Webster, R.~L., \& Turner, E.~L.\ 1999, \mnras, 309, 261

\bibitem[Wyithe et al.(2000)]{Wyithe00} Wyithe, J.~S.~B., Webster, R.~L., \& Turner, E.~L.\ 2000, \mnras, 315, 51 

\bibitem[Wyithe et al.(2000b)]{Wyithe00b} Wyithe, J.~S.~B., Webster, R.~L., \& Turner, E.~L.\ 2000, \mnras, 318, 762

\bibitem[Yee \& De Robertis(1991)]{Yee91} Yee, H.K.C. \& De Robertis, M.M. 1991, \apj, 381, 386

\bibitem[Yonehara(2001)]{Yonehara01} Yonehara, A.\ 2001, \apjl, 548, L127

\bibitem[Yoo et al.(2005)]{Yoo05} Yoo, J., Kochanek, C.~S., Falco, E.~E., \& McLeod, B.~A.\ 2005, \apj, 626, 51

\bibitem[Yoo et al.(2006)]{Yoo06} Yoo, J., Kochanek, C.~S., Falco, E.~E., \& McLeod, B.~A.\ 2006, \apj, 642, 22

\end{thebibliography}
\end{document}